# A Computational Approach to Historical Ontologies

Mat Kelly*, Jane Greenberg*, Christopher B. Rauch*, Sam Grabus*, Joan P. Boone*,
John A. Kunze†, and Peter Melville Logan‡

*College of Computing and Informatics
Drexel University
Philadelphia, PA, USA
{mkelly, jg3243, cr625, smg383, jpb357}@drexel.edu

†California Digital Library
University of California
Oakland, CA, USA
jak@ucop.edu

‡Temple Libraries
Temple University
Philadelphia, PA, USA
perter.logan@temple.edu

*Abstract*—This paper presents a use case exploring the application of the Archival Resource Key (ARK) persistent identifier for promoting and maintaining ontologies. In particular we look at improving computation with an in-house ontology server in the context of temporally aligned vocabularies. This effort demonstrates the utility of ARKs in preparing historical ontologies for computational archival science.

*Index Terms*—Archival computation, archival science, historical ontologies

## I. Introduction

Over the last 30 years, governmental research agencies, foundations, and institutions have collectively allocated millions of dollars toward digitization of nearly every type of archival artifact, regardless of genre or format. A necessary part of these processes includes the creation and integration of metadata to support resource discovery, access, and use. As these collections amount to big data, there is increased attention toward computational approaches with metadata in the context of computational archival science [1], [2]. Computation-ready metadata, increasingly referred to as smart and big metadata, can help contextualize an archival collection and offer new insights into the resources being described [3], [4]. Unfortunately, the full exploitation of topical or subject-oriented metadata is hampered by the absence of temporally relevant ontologies and the challenges with persistence among ontological resources terminologies.

Large scale digital archival collections simply use contemporary ontologies such as the Library of Congress Subject Headings (LCSH), one of the most expansive, general domain terminologies available in linked data. A key limitation is that contemporary ontologies, despite being computation-ready, are at odds with providing contextual interpretation and analysis of historical archival resources. Researchers may use word embeddings and data mining to understand historical concepts and the context of a particular collection [5]; however, this approach is not integrated with ontology standards and lacks the semantic infrastructure that can help support the study of concept and context over time. Research applying ontologies to historical records requires a more holistic approach, one that can be embedded in the practice of computational archival research. Specifically, there is a need for computation-ready ontologies that temporally align with the resources being represented. Research in this area also needs to consider persistent identifiers (long-term, stable data links) for ontological concepts to support longitudinal analysis of digital archival artifacts over time.

Fig. 1. A catalog entry for the LCSH1910 vocabulary.

Initial work in this area has been pursued as part of the 19th Century Knowledge project, where we have converted historical ontologies, specifically the 1750 classification system underlying Ephraim Chambers Cyclopaedia [6] and the 1910 version of the general domain, Library of Congress Subject Headings (1910 LCSH) into linked-data. With regard to scale for the dereferenced resource, the 1910 Library of Congress Subject Headings contain over 29,000 entries used in the dictionary catalogs of the Library of Congress published in 1910–1928. Figure 1 illustrates a sample catalog entry for the Armories entry. The use of these historical ontologies provides insight into historical collections in ways not possible with contemporary systems [7], [8]. A critical goal underlying the work, and the focus of this paper, is to establish resolvable, Persistent Identifiers (PIDs) for historical ontologies used for archives.

This paper presents our initiative exhibiting a preliminary use case with Archival Resource Keys (ARKs) as PIDs for the 1910 LCSH terms. The use of PIDs is crucial for the application and sharing of historical ontologies across diverse digital archives and, ultimately, for supporting deeper and more extensive computational archival science. We support this specific effort by integrating ARKs into the Helping Interdisciplinary Vocabulary Engineering (HIVE) ontology server. The next section provides an overview of preliminary work with historical ontologies and their relevance with computational archival science, followed by the case for ARKs. We then present the goals of our work and provide details on our methods. Next, we present our initial implementation exhibiting the approach to serve as the basis for further evaluation beyond this preliminary study. The paper wraps up with a contextual discussion of our work to date, and the conclusion summarizes current progress and identifies next steps.

TABLE I
1910 LCSH INDEXING RESULTS FOR THE 7TH ED. ENCYCLOPEDIA
BRITANNICA ENTRY ON SLAVERY AND MARK TWAIN'S "LETTER TO
PAMELA A. MOFFETT ON SEPTEMBER 4, 1874."

| Digital Archival Resource | Terms exclusive to 1910 LCSH vocabulary: |
|---|---|
| 7th Edition Encyclopedia Britannica Article on "Slavery" (19 C. Knowledge Project) | Houses; Carries; **Man**; **Negroes**; War; Government; Accounting; Accountants; Law; Societies; Age |
| Letter to Pamela A. Moffett on September 4, 1874 (Mark Twain Collection) | **Idiots**; **Imbecility**; **Turning**; Lawyers; Commons; Schools; **School**; Lays; **Fall**; Judges; Asylums; Building |

## II. ONTOLOGICAL APPLICATIONS

As more digital archival projects are made accessible for scholarship, the use of historical ontologies for describing these resources are scarce. One exception is research with the Helping Interdisciplinary Vocabulary Engineering (HIVE) application and the Nineteenth-Century Knowledge Project (19th C. Knowledge Project) [8].

The 19th C. Knowledge Project is building an extensive, open, digital archival collection to support the study of knowledge and its transformation. Four historical editions of the Encyclopedia Britannica (3rd, 7th, 9th, and 11th editions), spanning 1797 to 1911, have been digitized for computational study. The combined data set includes over 100,000 encyclopedia entries, covering a broad range of topics. The project uses both the contemporary LCSH and the 1910 LCSH for topical representation. The use of both ontologies provides more comprehensive access to the material and facilitates study of knowledge over time. Table I illustrates this difference for an encyclopedia entry and a letter drawn from the Mark Twain Collection [9], where the terms in boldface indicate entries that no longer exist in the contemporary LCSH vocabulary as an exact match, indicating concept drift in the controlled vocabulary over time.

More recent research has compared the 1910 LCSH with OCLC's faceted rendering of the contemporary, 2020 LCSH, known as FAST [10]. A sample of 90 full text entries, drawn across the four historical editions of the Encyclopedia Britannica, through the examination of 886 indexing results, found that the 31.3% terms generated with the 1910 LCSH did not appear in the contemporary FAST automatic indexing results. Further analysis found that 6.2% of the total 1910 LCSH results no longer appear in the full 2020 FAST vocabulary, representing temporal concept drift.

This initial research confirms that computation-ready historical ontologies can provide insight into the archival record with access to language absent from contemporary systems. Additionally, a computation-ready historical ontology has applicability well beyond the 19th C. Knowledge Project and enables comparison of collections as demonstrated with the Twain Project letter. The conversion to linked data is an important first step but significantly limited without the support of reliable persistent identifiers supporting provenance tracking and preservation. The significance of persistent identifiers for doing this work has been emphasized in the life sciences [11]. This need underscores the case for computational archival science in the context of ontologies and informs our work adopting the ARK identifiers.

### A. Archival Resource Keys (ARKs)

ARKs (Archival Resource Keys) are high-functioning identifiers that provide references and descriptions for objects of any type [12]. The ARK specification is in active development. Generally, an ARK is a special kind of URI that connects users to the named object, its metadata, and the ARK service provider's promise about its persistence [13]. An ARK is high functioning in the sense that it provides access to not only the object and its metadata, but a specific, revealed commitment statement from the institutional provider responsible for the resource relating to access permanence.

*1) Persistence:* Impersistence is a major problem for linked data, with the average URL remaining valid (not returning a 404 Page Not Found error) for only 44 days [14]. In response, a number of PID schemes have been proposed (ARK, DOI, Handle, PURL, URN). While there is no guarantee that the payload to which any PID points is currently, ever was, and/or will remain accessible [15], a distinguishing characteristic of the ARK scheme is to provide a standard way to link from an object to a promise of stewardship for it. Combined with traditional reputation mechanisms, the details of that promise give link recipients the best basis on which to forecast persistence.

*2) Format:* An ARK identifier is carried in a Uniform Resource Identifier (URI) and shares some of its structure. As such it begins with the Name Mapping Authority (or hostname — see Figure 2) that would provide name resolution service (resolver) for the ARK, however, the core immutable identity of the ARK is actually independent of the hostname.

The Base Object Name (Figure 2) is composed of the label to signify an ARK (`ark:/`), a Name Assigned Authority Number (NAAN, e.g., 99152), and an Assigned Name (e.g., 5p30086k). The character "/" after that name indicates containment. For example, b41910/5p30086k asserts that the object 5p30086k is contained by the object b41910. While any alphanumeric string can be used for the Assigned Name, the specification recommends using a "betanumeric" [12] character set which removes all vowels and the lowercase l (letter ell) to mitigate ambiguity and help limit transcription errors. More than 667 registered organizations have created an estimated 8.2 billion ARKs [16].

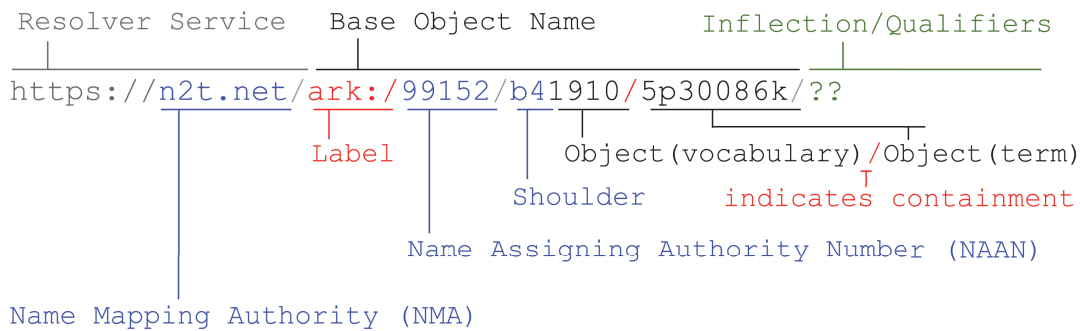

Fig. 2. Anatomy of a URI containing an ARK

A Name Assigning Authority Number (NAAN) is a unique identifier reserved for an organization that wants to create ARKs. Prepending an organization's NAAN to its Assigned Names is required to keep the organization's ARKs globally unique. The ARK also supports shared NAANs to permit leveraging certain specific immutable semantics: 12345 for example ARKs (such as those appearing in documentation); 99152 for metadata, controlled vocabulary, and ontology terms; 99166 for agents (such as people, groups, and institutions); and 99999 for testing and development. We use the 99152 NAAN in this work so that receiving software can immediately infer that the identified object is a "term."

To use a shared NAAN without conflicts requires reserving a "shoulder" (a term borrowed from a part of a physical, metal key), or sub-namespace under a NAAN [17]. Figure 2 displays these and other components of an ARK encoded as a URI to be dereferenced with a resolver. By convention, shoulders represent a short, fixed extension to the NAAN. For example `ark:/99152/b4`.

*3) What's Different About ARKs:* Most ARKs are created by organizations that tend to publish them based on their own resolvers, but ARKs are intended to be decoupled from any particular resolver. For example, the ARK `ark:/12345/x54xz321` might be resolved with the Name-to-Thing resolver (currently administered by the California Digital Library [17]) at the online hostname n2t.net and dereferenced at https://n2t.net/ark:/12345/x54xz321. However, the concept of providing the identifier without a hostname (i.e., `ark:/12345/x54xz321`) is an expected usage pattern that allows the identifier to persist and be supported by successor or even parallel name mapping authorities.

ARKs are atypical compared to other PIDs in that there are no fees associated with obtaining a NAAN and publishing ARKs, the options to specify metadata are flexible by design, and resolution is decentralized, despite a recommended but not required resolver (N2T) [15]. This is in contrast to other PIDs like DOIs [18], Handles [19], PURLs [20], and URNs [21], all of which have requirements for centralization, rigidity with respect to metadata requirements, cost, etc. In addition, all of these schemes ultimately have demonstrated little effect on persistence [13]. A founding principle of the ARK scheme is that persistence is a matter of service and that there must be a way to tie a promise of persistence to a provider's reputation in that area. Persistence is achieved through a provider's successful stewardship of objects and their identifiers [20].

*B. Helping Interdisciplinary Vocabulary Engineering (HIVE)*

HIVE is a linked data, automatic indexing metadata application (Figure 3). HIVE enables researchers to leverage existing ontologies and automatically generate well-formed, standardized metadata for digital resources. One of the key features of HIVE is that it supports the use of multiple ontologies during a single metadata generation sequence. A curator can work with both historical and contemporary ontologies in a single automatic indexing sequence while also including ontologies representing different domains. For instance, a curator may seek to use both the contemporary LCSH and a geospatial ontology [22].

HIVE's indexing algorithm identifies candidate keyphrases in digital sources and then compares them to one or more user-selected ontologies in order to suggest a set of terms for content representation. The HIVE system supports curators in ultimately selecting the best ontology terms, although the output can also be automatically assigned to digital content, an approach being pursued with the 19th Century Knowledge Project.

Integration of the ARK Identifier schema to historical ontologies can address such shortcomings and improve discovery and access to cultural heritage resources. The attribution of reputation related metadata may help to mitigate the current state of affairs where no system of authority files for linking data to historical ontologies exists.

III. METHODOLOGY

As directed by the ARK FAQ, we registered a NAAN (13183) for the Metadata Research Center at Drexel University (MRC) and a shoulder (b4) to designate vocabularies on the existing shared NAAN (99152). A primary focus of the work described in this paper was to resolve identifiers under our NAAN (13183), supplemented with ARKs specifically for ontology terms using the 99152/b4 shoulder.

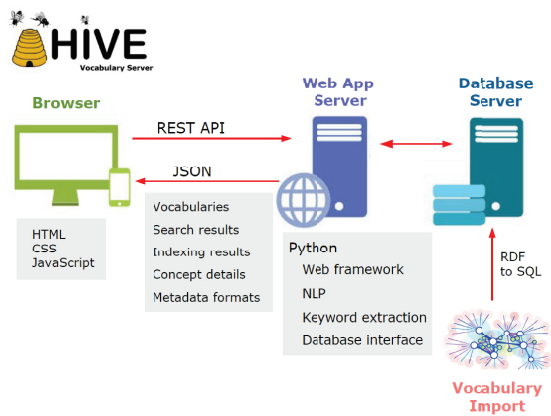

Fig. 3. The HIVE Vocabulary Server implements a client-server, web-based interface to provide access to a selection of vocabularies and linked data.

After having reserved our organization's NAAN, we considered multiple variations of path structures to plan for extensibility beyond the use case we describe in this paper. In our initially considered variation, we thought to encode the vocabulary name into the Assigned Name, e.g., `ark:/13183/lcsh1910/5p30086k` (where lcsh1910 is the name of the vocabulary). This configuration, however, does not follow ARK best practices ("betanumeric" characters and "first-digit convention" as per the Shoulder FAQ [16].

A second structure we considered was to assign one name to each vocabulary under the NAAN, e.g., b412345. This would produce ARKs like https://n2t.net/ark:/13183/b412345/lcsh1910/5p30086k. This configuration incorrectly implies that b412345 names an actual object with sub-objects.

Ultimately, we settled on the configuration where the portion of the ARK indicating the lcsh1910 vocabulary is not semantically encoded but still represented uniquely. Rather than b412345 representing all vocabularies, the b4 shoulder becomes the root of a namespace and an appended substring (e.g., 1910) indicates the LCSH 1910 vocabulary.

The choice to use "b41910" instead of the more human-readable "lcsh" is more opaque (widely seen to benefit persistence), more textually efficient than the alternatives we considered, leverages a shoulder for future vocabularies, and allows the NAAN to be reused for ARKs that refer to other things by changing the shoulder. A downside of the b41910 assigned name is that other vocabularies might not have temporal semantics (like the LCSH vocabulary using 1910), thus introducing a schematic discrepancy between vocabularies. Also, the numerical value still encodes some semantics, i.e., suboptimal opacity.

## IV. Implementation

As an initial prototype and proof-of-concept for the approach, we sought to leverage the advantages of the ARK ecosystem, publish the vocabularies online, and connect the identifiers to the vocabulary payload to make them accessible. One ultimate goal of this process is to make this vocabulary accessible online to the software implementation of HIVE. In this section, we provide details on this process.

### A. Establishing PIDs

Our initial approach in establishing ARKs to serve as persistent identifiers for vocabularies takes into account scalability and extensibility with an emphasis on persistence. The ongoing process is progressing from both ends to meet in the middle, i.e., establish PIDs, progress toward resolution, strategically publish the payload (vocabularies), and work towards a reliable means of access.

The Noid (Nice Opaque IDentifier) utility can mint (generate) transcription safe, unique strings that adhere to the preferred betanumeric recommendations for ARKs [23] supported by the N2T.net resolver and the ARK standard. The tool and/or service is used by a variety of institutions inclusive of the Internet Archive (NAAN 13960), University of California Berkeley (28722), National Library of France (12148), Portico Digital Preservation Service (27927), and Smithsonian Institution (65665). A comprehensive, updated list of NAANs is available online [24].

Resolving our vocabulary entries involves a URI redirect chain for which an ARK acts as the common thread. As an example, upon being provided an ARK like `ark:/99152/b41910/5p30086k`, a client or system (henceforth "user") wanting to leverage the vocabulary could resolve the ARK. Through MRC being responsible for the shoulder b4 under NAAN 99152, this URI directs the user to https://id.cci.drexel.edu/lcsh1910/5p30086k for further routing.

### B. Publishing the Vocabulary

As an additional assurance of persistence that forms one of the essential elements of a provider's commitment to stewardship, this URI is not redirected using HTTP but rather mapped using DNS to GitHub Pages, the contents of which reside in the repository in https://github.com/metadata-research/vocabularies/lcsh1910. The opaque ARK identifier would ultimately redirect to https://github.com/metadata-research/vocabularies/lcsh1910/5p30086k for the period of time that GitHub hosting is used, and would be updated in the future when the hosting arrangement changed. MRC is responsible for the response to any requests as forwarded from a resolver for the shoulder b4 under the NAAN 99152. This design choice helps to ensure data accessibility beyond the confines of a single provider while avoiding possible user misconstruction of the ARK identifier. Further, hosting the vocabularies on GitHub encourages version tracking of any vocabularies that are still progressively being generated (e.g., on contemporary topics) or refined (for historical ontologies from perhaps not born digital sources) to facilitate accountability, openness, and community engagement.

*C. Incorporating ARKs into HIVE*

The current implementation of HIVE defines a unique URL for each term in the 1910 LCSH. Our initial prototypes and mock-ups use the shared NAAN for display purposes. These URLs will be replaced with unique and opaque ARK identifiers. The Noid utility is used to generate (mint) a unique core identifier that includes the NAAN and shoulder value. This identifier is then embedded in a URL to form the ARK identifier, for example, `ark:/99152/b41910/5p30086k`. The 1910 LCSH database in HIVE will then be regenerated to incorporate ARK identifiers for each vocabulary term.

Figure 4 illustrates the HIVE web page for the Abbeys entry in the 1910 LCSH. The value for the URI/URL has been replaced with a URI/PID field containing a unique ARK identifier. The "Related" entries (Cathedrals, Convents, Monasteries) will also be assigned ARK identifiers and linked to their respective terms.

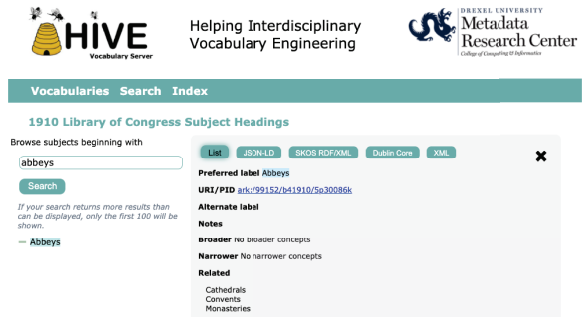

Fig. 4. A catalog entry for the LCSH1910 vocabulary.

## V. Discussion

The case we present in this paper and our approach of uniquely identifying a historical vocabulary using the ARK standard and subsequently, an ontology exploration tool, was a first, exploratory step in determining the usefulness, feasibility, and merit of making the data available using emerging standards. The ARK ecosystem is evolving (i.e., currently an IETF standards draft [25]) and thus exploration of this sort can serve as a basis for practice by others that would like to accomplish similar goals.

While using the Git distributed versioning protocol (via GitHub) as a repository for data storage does not ensure permanence, we are considering additional strategies for redundancy. However, the integration into a system that inherently instills conventional distributed version control may provide additional use for vocabularies that are still evolving, unlike the static LCSH1910. Contrarily, the public availability of the data might further facilitate use and reuse of the data itself beyond implementations we can initially anticipate in this paper.

There is an apparent nascent interest in historical ontologies, such as the use of multiple period vocabularies to contextualize nineteenth-century sideshow performer images in the Ronald G. Becker Collection of Charles Esenmann Photographs [26].

Another example is the PeriodO Project, a collaboration between University of Austin, Texas and University of North Carolina, which has explored the availability of historical terms for representing time-periods, using its own "shoulder", `ark:/99152/p0`, to represent historical concepts as linked data [27]. Advancing towards a more extensive environment in which entire historical ontologies can be located and applied to archival collections, concepts need to be resolved using persistent identifiers such as ARKs.

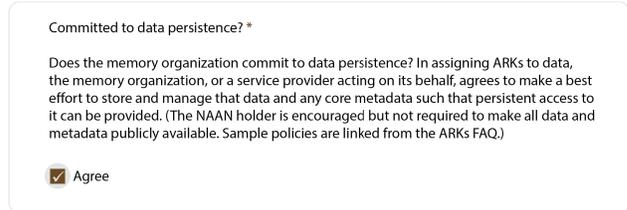

Fig. 5. Registering a NAAN requires the organization to commit to data persistence by checking a box.

Persistence as a chief motivator for this work is coupled with commitment to stewardship. At this time, persistence in any PID scheme cannot be enforced, and is usually approached by a combination of up-front declaration of intention (e.g., see Figure 5), broken link monitoring, reporting, and sometimes community intervention (e.g., the Internet Archive steps in to rescue content from an collection that has lost funding). Our initial plan to use Git is a step toward this and GitHub's commitment to ensuring long term preservation is demonstrated via the recent Arctic Code Vault endeavor. However, limitations surface when considering how often updates will be made as well as threats posed by climate change, where the future existence of the arctic is at risk.

While ARKs are protocol agnostic, the current expectation is for the resolver to be accessed via HTTP to dereference the ARK embedded within a URI. However, the scheme/protocol (qua RFC3986 [28]) is not embedded in the core immutable ARK, allowing the dereferencing procedure to potentially use another system beyond HTTP for ARK resolution. This aspect of portability allows ARKs to be relevant to systems beyond the Web. The initial work we investigated and performed in this paper has helped our team to more fully appreciate and understand the immense flexibility and durability of the ARK ecosystem and anticipate future use cases. This may prove significant for ontological systems to operate in other electronic and analog environments – critical for the persistence of historical vocabularies.

In manifesting the key archival principles of evidential and enduring value [29] and grounding the work presented here in the context of computational archival science (CAS), the value of ARKs for tracking provenance and promoting persistence are key. While this case study focuses on the 1910 LCSH, the work with ARKs has implications for a full range of historical ontologies, many of which remain in analog format. The implications of this research extend beyond LCSH and humanities resources, and demonstrate the potential for converting other historical ontologies into linked data, facilitated by ARK identifiers. This approach could potentially be applied to scientific, medical, agricultural, and other historical terminologies.

## VI. CONCLUSIONS

In this paper, we present a case exploring the appplication of ARK persistent identifiers for the 1910 Library of Congress Subject Headings by registering our own shoulder under the shared NAAN for terms and investigating an approach for establishing ARKs in a systematic way. Our initial prototypes and mock-ups in the context of the HIVE tool allow for further integration of other vocabularies and provide a scalable approach for integrating PIDs into HIVE in the future. Facilitating the availability of linked historical data through the efforts described in this paper is a first step in preparing historical ontologies for computational archival science in a wide variety of contexts.

## VII. FUTURE WORK

In our initial exploration of using ARKs for a historical ontology, we sought to facilitate persistence of this data by preventing reliance on a single service or data source. For the time being, the N2T resolution service is primarily responsible for resolving ARKs. In the future, we hope to establish a method and practice of propagating the resolution of ARKs to an alternative data source to encourage more distributed ARK resolution.

This initial effort was an exploration of the applicability and feasibility of utilizing an existing PID system as a means of resolving and facilitating access to the LCSH vocabulary. We exhibited the feasibility of systematically integrating ARKs with a historical vocabulary and anticipate refining our process. We also plan to apply our initial process to other relevant vocabularies.

This initial exploration proved valuable as a low-risk, preliminary exercise in the potential to integrate PIDs with historical vocabularies. We learned a lot about both systems through this exploration and realize that there are still many unknowns whose challenges will only be surfaced through exploration and prototyping of this nature. The long-term impact of making historical ontologies available with PIDs has implications not only for historical documents in the humanities but in enabling the computability of archives. There is more we can learn from archival practice in the areas of provenance, life-cycle management, use, and reuse that can be applied to big data.

## REFERENCES

[1] R. Marciano et al., "Reframing Digital Curation Practices through a Computational Thinking Framework," in *2019 IEEE International Conference on Big Data (Big Data)*, Dec. 2019, pp. 3126–3135.
[2] N. Payne, "Stirring The Cauldron: Redefining Computational Archival Science (CAS) For The Big Data Domain," *2018 IEEE International Conference on Big Data (Big Data)*, 2018.
[3] J. Chen, W. Lu, and O. Zavalina, "Workshop on Organizing Data, Information, and Knowledge in Big Data Environments," in *2019 ACM/IEEE Joint Conference on Digital Libraries (JCDL)*, Jun. 2019, pp. 459–460.
[4] M. L. Zeng, "Smart Data for Digital Humanities," *Journal of Data and Information Science*, vol. 2, no. 1, pp. 1–12, Feb. 2017.
[5] W. L. Hamilton, J. Leskovec, and D. Jurafsky, "Diachronic Word Embeddings Reveal Statistical Laws of Semantic Change," *arXiv:1605.09096 [cs]*, Oct. 2018. [Online]. Available: http://arxiv.org/abs/1605.09096
[6] E. Chambers, *Chambers' Cyclopaedia*, 1750. [Online]. Available: https://artfl-project.uchicago.edu/content/chambers-cyclopaedia
[7] S. Grabus, J. Greenberg, P. Logan, and J. Boone, "Representing Aboutness: Automatically Indexing 19th- Century Encyclopedia Britannica Entries," *NASKO*, vol. 7, no. 1, pp. 138–148, Sep. 2019, number: 1.
[8] P. M. Logan, "Nineteenth-Century Knowledge Project," 2019. [Online]. Available: https://tu-plogan.github.io/
[9] "Mark Twain Project." [Online]. Available: https://www.marktwainproject.org/
[10] L. M. Chan and E. T. O'Neill, *FAST: Faceted Application of Subject Terminology*. Libraries Unlimited, 2010. [Online]. Available: https://products.abc-clio.com/abc-cliocorporate/product.aspx?pc=F2275P
[11] J. A. McMurry et al., "Identifiers for the 21st century: How to design, provision, and reuse persistent identifiers to maximize utility and impact of life science data," *PLOS Biology*, vol. 15, no. 6, p. e2001414, Jun. 2017, publisher: Public Library of Science. [Online]. Available: https://journals.plos.org/plosbiology/article?id=10.1371/journal.pbio.2001414
[12] J. A. Kunze, "ARK Identifiers FAQ - ARKs in the Open - LYRASIS Wiki." [Online]. Available: https://wiki.lyrasis.org/display/ARKs/ARK+Identifiers+FAQ
[13] J. Kunze, "Towards Electronic Persistence Using ARK Identifiers," Aug. 2003. [Online]. Available: https://escholarship.org/uc/item/3bg2w3vs
[14] B. Kahle, "Preserving the Internet." [Online]. Available: https://www.scientificamerican.com/article/preserving-the-internet/
[15] J. Kunze, "Ten persistent myths about persistent identifiers," Aug. 2018. [Online]. Available: https://escholarship.org/uc/item/73m910w8
[16] J. A. Kunze, "ARK Shoulders FAQ - ARKs in the Open - LYRASIS Wiki." [Online]. Available: https://wiki.lyrasis.org/display/ARKs/ARK+Shoulders+FAQ
[17] "EZID: Identifier Concepts and Practices at the California Digital Library." [Online]. Available: https://ezid.cdlib.org/learn/id_concepts
[18] "ISO Information and documentation — Digital object identifier system 26324:2012." [Online]. Available: https://www.iso.org/cms/render/live/en/sites/isoorg/contents/data/standard/04/35/43506.html
[19] S. Sun, L. Lannom, and B. Boesch, "Handle System Overview." [Online]. Available: https://www.rfc-editor.org/rfc/rfc3650
[20] K. E. Shafer, "ARMs, OCLC Internet Services, and PURLs," *Journal of Library Administration*, vol. 34, no. 3-4, pp. 387–393, Dec. 2001, publisher: Routledge _eprint: https://doi.org/10.1300/J111v34n03_19. [Online]. Available: https://doi.org/10.1300/J111v34n03_19
[21] P. Saint-Andre and J. C. Klensin, "Uniform Resource Names (URNs)," Internet Requests for Comments, RFC Editor, RFC 8141, April 2017. [Online]. Available: https://www.rfc-editor.org/rfc/rfc8141.txt
[22] Y. Zhang, A. Ogletree, J. Greenberg, and C. Rowell, "Controlled vocabularies for scientific data: Users and desired functionalities," *Proceedings of the Association for Information Science and Technology*, vol. 52, no. 1, pp. 1–8, 2015. [Online]. Available: https://asistdl.onlinelibrary.wiley.com/
[23] "NOID." [Online]. Available: https://n2t.net/e/noid.html
[24] University of California, California Digital Library (CDL), "Name Assigning Authority Number (NAAN) Registry." [Online]. Available: https://n2t.net/e/pub/naan_registry.txt
[25] J. A. Kunze, "The ARK Identifier Scheme Draft 24." [Online]. Available: https://datatracker.ietf.org/doc/draft-kunze-ark/
[26] B. Dobreski, J. Qin, and M. Resnick, "Side by Side: The Use of Multiple Subject Languages in Capturing Shifting Contexts around Historical Collections," *NASKO*, vol. 7, no. 1, pp. 16–26, Sep. 2019, number: 1.
[27] A. Rabinowitz, "It's About Time: Historical Periodization and Linked Ancient World Data." [Online]. Available: http://dlib.nyu.edu/awdl/isaw/isaw-papers/7/rabinowitz/
[28] T. Berners-Lee, R. T. Fielding, and R. L. Masinter, "Uniform Resource Identifier (URI): Generic Syntax," Internet Requests for Comments, RFC Editor, RFC 3986, January 2005. [Online]. Available: https://www.rfc-editor.org/rfc/rfc3986.txt
[29] T. R. Schellenberg, *Modern archives: principles and techniques*. University of Chicago Press, 1956.